\documentclass[seceq]{ptptex}

\usepackage{graphicx}
\usepackage{hyperref}

\title{
Similarity and Probability Distribution Functions in Many-body Stochastic Processes with Multiplicative Interactions
}

\author{
Akihiro {\sc Fujihara}$^1$\footnote{email: fujihara@yokohama-cu.ac.jp}, Toshiya {\sc Ohtsuki}$^2$, and Hiroshi {\sc Yamamoto}$^2$
}

\inst{
$^1$Graduate School of Integrated Science, Yokohama City University, 22-2 Seto, Kanazawa-ku, Yokohama 236-0027, Japan\\
$^2$Field of Natural Science, International Graduate School of Arts and Sciences, Yokohama City University, 22-2 Seto, Kanazawa-ku, Yokohama 236-0027, Japan
}

\recdate{
\today
}

\abst{
 Analytical and numerical studies on many-body stochastic processes with multiplicative interactions are reviewed. The method of moment relations is used to investigate effects of asymmetry and randomness in interactions. Probability distribution functions of the processes generally have similarity solutions with power-law tails. Growth rates of the system and power-law exponents of the tails are determined via transcendental equations. Good agreement is achieved between analytical calculations and Monte Carlo simulations. 
}

\begin{document}

\maketitle

\section{Introduction}

 For several decates, power-law distributions have been one of the central issues in statistical physics and have been observed not only in physical, but also in biological, social, and economic phenomena. In order to explain this spontaneous emergence of power laws, several concepts, such as self-organized criticality\cite{rf:BTW1987, rf:Jensen1998, rf:Sornette2000}, scale-free network\cite{rf:AB2002}, and so on, have been proposed. Some power laws have been successfully explained by the concepts, but others are not sufficiently. A lot of researchers have strenuously engaged in finding underlying physics of the power-law distributions. 

 In recent years, inelastic Maxwell models dealing with the system of inelastically colliding particles have been investigated and have revealed that the velocity distribution function of particles has a power-law tail\cite{rf:EB2002, rf:BK2002}. Inelastic Maxwell model(IMM) is defined on $N-$particle system. Each particle has a positive quantity $x_i$\;($i=1, \ldots, N$) and at each timestep, two of them are selected randomly and experience a binary interaction for transforming the two quantities of particles $x_i, x_j$ into $x_i' = \epsilon x_i + (1 - \epsilon) x_j, \ x_j' = (1 - \epsilon) x_i + \epsilon x_j$ where $0 < \epsilon < 1$ is a restitution coefficient. Similar models such as greedy mutiplicative exchange(GME)\cite{rf:IKR1998} whose interaction is described by $x_i' = (1 - \alpha) x_i, \ x_j' = x_j + \alpha x_i \; (x_j > x_i)$ where $0 < \alpha < 1$ is a interaction parameter and symmetrically multiplicative interaction(SMI)\cite{rf:bBLR2003} whose interaction is $x_i' = \alpha x_i + \beta x_j, \ x_j' = \beta x_i + \alpha x_j$ where $\alpha, \beta > 0$ have also been studied. Characteristic features common to these models are that they have similarity solutions scaled by a growth rate of the system and power-law tails in the probability distribution function(PDF) of the quantities. In general, the power-law tails of these models appear if their interaction parameters are within a finite \textit{region} rather than a critical \textit{point}. This means that fine-tuning of the parameters is not needed for observing the power laws, while it is necessary in usual critical phenomena. In addition, it is also found that the power-law exponents, which are able to be calculated analytically, vary continuously with the values of interaction parameters. 

 In this paper, we collectively call these models many-body stochastic processes with multiplicative interactions and consider two extended models: asymmetrically multiplicative interaction and effects of randomness in interaction. The contents are organized as follows. Firstly, key points of theoretical procedures of analyzing the processes are briefly introduced. Secondly, a model of asymmetrically multiplicative interaction is considered. Thirdly, a model which takes into account effects of randomness in interaction is considered. Finally, we conclude with a summary and application of our models.

\section{Key points of the theory of many-body stochastic processes with multiplicative interactions}\label{sec:keypoint}

 Let us introduce general procedures to investigate the tails of PDFs of the processes analytically. The method of Fourier transform has been often adopted for this issue\cite{rf:EB2002, rf:BK2002, rf:bBLR2003, rf:OFY2004, rf:FOY2004, rf:Slanina2004}. However, we would like to consider it here with using a method of moment relations which is easier to understand it intuitively. We will apply this method to our models in the next two sections. In order to introduce the procedures, let us take SMI whose interaction rule is $x' = \alpha x + \beta y, \ y' = \beta x + \alpha y$ as an example. In the limit the number of particles $N \to \infty$, a master equation of the system is descirbed as follows. 
\begin{equation}
\frac{\partial }{\partial t} f(z, t) = \int_{0}^{\infty}dx \int_{0}^{\infty}dy\ f(x, t)\ f(y, t)\ \left[ \delta(z - (\alpha x + \beta y)) - \delta(z - x) \right]. \label{eq:key_master}
\end{equation}
The goal of our calculations is to find a similarity solution of this equation. Thus, the following scaling relations are assumed. 
\begin{equation}
\xi = z\ e^{- \gamma t}, \hspace{5mm} f(z, t) = e^{- \gamma t}\ \Psi (\xi), \label{eq:scaling}
\end{equation}
where $\gamma$ which is obtained explicitly later is a growth rate of the system. Substituting Eq.~(\ref{eq:scaling}) into Eq.~(\ref{eq:key_master}) and some calculations with using Lemma 2 in Bobylev {\it et al.}'s paper\cite{rf:BGP2003}, the next inequalities of moments are obtained. 
\vspace{-2mm}
\begin{eqnarray}
\frac{\sum_{k=1}^{k_p-1}\left(\begin{array}{c}p\\k\end{array}\right)(\alpha^{k}\beta^{p-k}+\alpha^{p-k}\beta^{k}) \mu_{k} \mu_{p-k} }{ p \gamma - ( \alpha^{p} + \beta^{p} - 1 ) } \hspace{10mm} \nonumber \\
 \hspace{10mm} \le \ \mu_{p} \ \le \ \frac{\sum_{k=1}^{k_p}\left(\begin{array}{c}p\\k\end{array}\right)(\alpha^{k}\beta^{p-k}+\alpha^{p-k}\beta^{k}) \mu_{k} \mu_{p-k} }{ p \gamma - ( \alpha^{p} + \beta^{p} - 1 ) }, \label{eq:key_inequalities}
\end{eqnarray}
where $\mu_p$ is the $p-$th order moment of $\Psi(\xi)$ and $k_p$ denotes the integer part of $(p+1)/2$ for $p \ge 1$. Note that when $p$ is an odd integer, the first inequality becomes an equality. This determines the growth rate $\gamma = \alpha + \beta - 1$ at $p=1$. The moment inequalities (\ref{eq:key_inequalities}) indicate that if a transcendental equation $p \gamma = \alpha^p + \beta^p - 1$ has a solution except a trivial one $p=1$, let us say $p=s$, then the moments $\mu_p$ of $(1 <)\ p < s$ are bounded and those of $p \ge s \ (> 1)$ diverge. This immediately lead to a power-law tail with the exponent $s$ in PDF $\Psi(\xi) \sim 1 / \xi^{1+s}$ where $\xi \gg 1$. When $0 \le s \le 1$, this formalism based on Eq.~(\ref{eq:key_inequalities}) breaks down. Instead, an inequality $p \gamma \le \alpha^p + \beta^p - 1$ can be derived in this case. This indicates that a power-law tail appears if the system is scaled by the minimum growth rate $\gamma_{min}$ under the condition that $s \gamma_{min} = \alpha^s + \beta^s - 1$ has a solution. Thus, the exponent $s$ is determined by $\{ \alpha^s \ln \alpha + \beta^s \ln \beta \} s = \alpha^s + \beta^s - 1$. 

 Consequently, the power-law exponent $s$ is determined by transcendental equations $(\alpha + \beta - 1) s = \alpha^s + \beta^s - 1 \ (s > 1)$ and $\{ \alpha^s \ln \alpha + \beta^s \ln \beta \} s = \alpha^s + \beta^s - 1 \ (0 \le s \le 1)$ and the growth rate becomes $\gamma = \alpha + \beta - 1 \ (s > 1)$ and $\gamma = \gamma_{min} \ (0 \le s \le 1)$. This denotes that $\gamma$ and $s$ are continuous functions of interaction parameters $\alpha$ and $\beta$. These are the key points of the theory of many-body stochastic processes with multiplicative interactions.

\section{Effects of asymmetry in interaction}

\begin{figure}
\begin{tabular}{cc}
\resizebox{60mm}{!}{\includegraphics{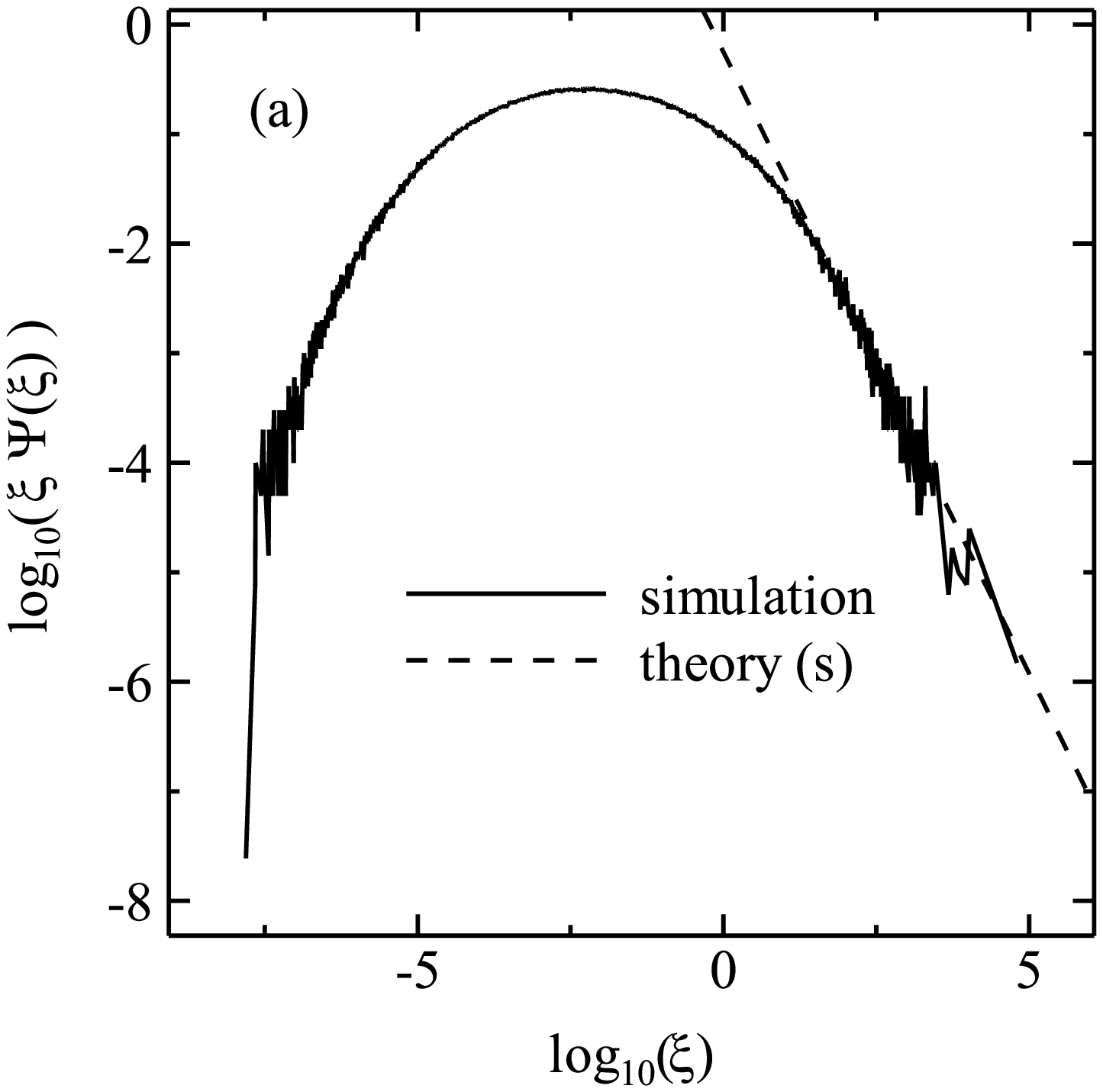}}
\resizebox{60mm}{!}{\includegraphics{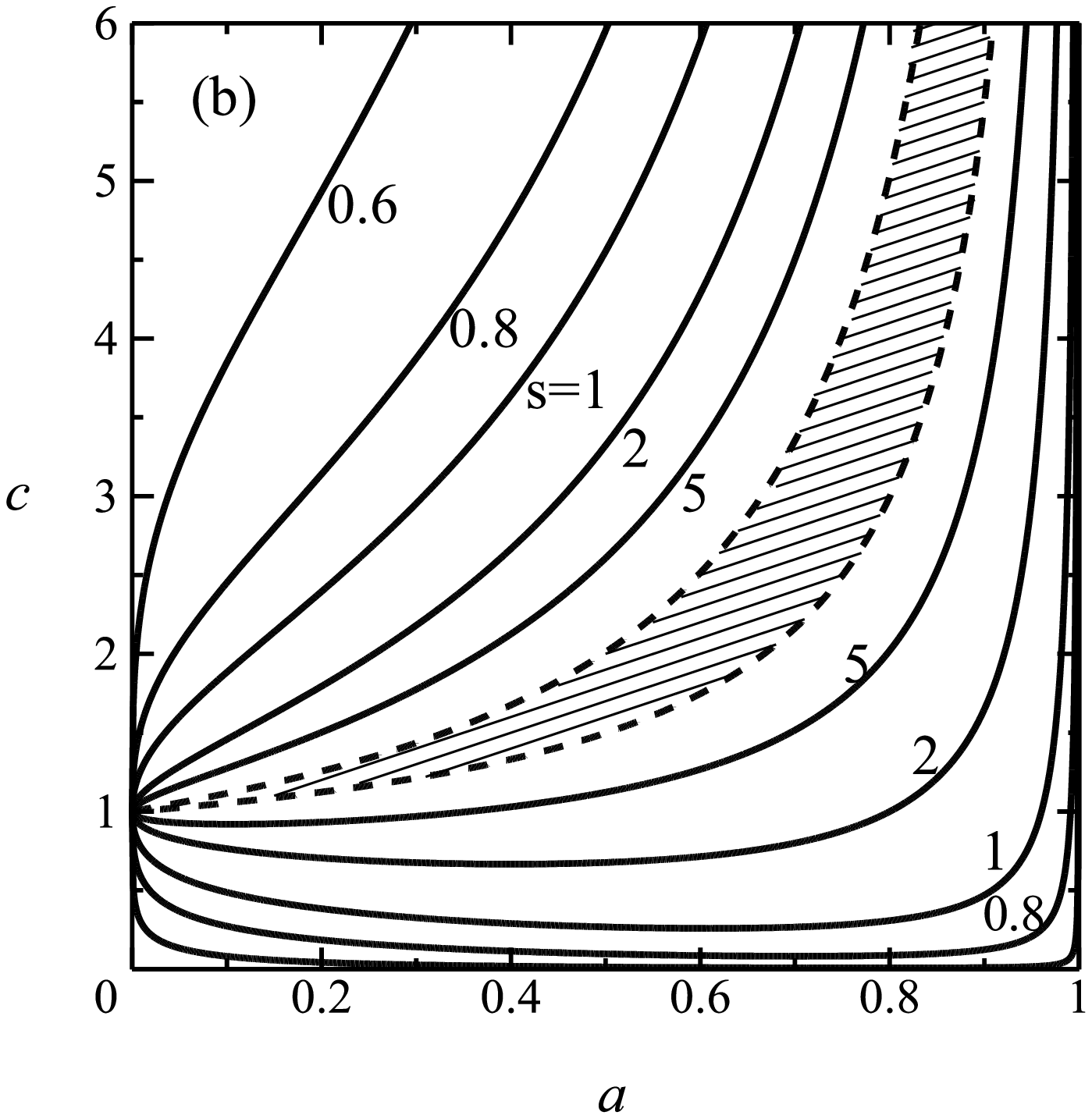}}
\end{tabular}
\caption{(a) Double logarithmic plot of $\xi \Psi(\xi)$ versus $\xi$ at $a=0.05, b=0.9, c=1.5, d=0.5$. Number of particles in a Monte Carlo simulation is $N=10^6$ and number of interactions is $T=50N$. (b) Phase diagram of the exponent $s$ in the case $a+b=1$ at $d=0.5$. Solid curves denote contours of $s$. Power-law tails disappear in the hatched region surrounded by dashed curves.}
\label{fig:asymmetry}
\end{figure}

 In this section, we investigate many-body stochastic processes with asymmetrically multiplicative interactions(AMI)\cite{rf:FOY2004}. A interaction rule of the processes is expressed as $x' = c(1-a) x + cb y, \ y' = da x + d(1-b) y \ (x \ge y)$ where $0 \le a, b \le 1$ and $c, d > 0$ are interaction parameters representing exchange rates and amplification rates of larger and smaller quantities, respectively. This rule distinguishes one interacting particle from the other by the magnitude of thier quantities $x$, $y$. This difference of interaction rule between two interacting particles is what we call asymmetry. Note that this model includes the three models IMM($c=d=1, \ a=b$), GMI($c=d=1, \ a=0$), and SMI($c=d, \ a=b$) aforesaid as special cases. We have performed Monte Carlo simulations of this model and have found that PDF obeys a power law at the tail as shown in Fig.~\ref{fig:asymmetry}(a). So, we examine this power-law tail in accordance with the procedures mentioned in Sec.~\ref{sec:keypoint}. A master equation of this system is described by 
\begin{eqnarray}
\frac{\partial f(z,t)}{\partial t} + f(z,t) &=& \int_{0}^{\infty}dy \int_{y}^{\infty}dx f(x,t) f(y,t) \nonumber\\
					    & & \times \left[ \delta(z-(c(1-a)x+cby)) + \delta(z-(dax+d(1-b)y)) \right]. \label{eq:asym_mastereq}
\end{eqnarray}
 The same scaling relations (\ref{eq:scaling}) is also used in this case. Substituing Eq.~(\ref{eq:scaling}) into Eq.~(\ref{eq:asym_mastereq}) and some calculations lead to the following moment inequalities. 
\begin{eqnarray}
 \sum_{k=1}^{k_p-1} \left(\begin{array}{c}p\\k\end{array}\right) \left[ \right. \{c(1-a)\}^{k} \{cb\}^{p-k} + \{c(1-a)\}^{p-k} \{cb\}^{k} + \{da\}^{k} \{d(1-b)\}^{p-k} \nonumber \\
 + \{da\}^{p-k} \{d(1-b)\}^{k} \left. \right] \times \int_{0}^{\infty} d \xi_{2} \int_{\xi_{2}}^{\infty} d \xi_{1} (\xi_{1}^{k} \xi_{2}^{p-k} + \xi_{1}^{p-k} \xi_{2}^{k} ) \Psi(\xi_{1}) \Psi(\xi_{2}) \nonumber \\
 \le \ (p \gamma + 1) \mu_{p} - ( \{c(1-a)\}^{p} + \{da\}^{p} ) \int_{0}^{\infty} d \xi_{2} \int_{\xi_{2}}^{\infty} d \xi_{1} \xi_{1}^{p} \Psi(\xi_{1}) \Psi(\xi_{2}) \nonumber \\
 - ( \{cb\}^{p} + \{d(1-b)\}^{p} ) \int_{0}^{\infty} d \xi_{2} \int_{\xi_{2}}^{\infty} d \xi_{1} \xi_{2}^{p} \Psi(\xi_{1}) \Psi(\xi_{2}) \nonumber \\
 \le \ \sum_{k=1}^{k_p}\left(\begin{array}{c}p\\k\end{array}\right) \left[ \right. \{c(1-a)\}^{k} \{cb\}^{p-k} + \{c(1-a)\}^{p-k} \{cb\}^{k} + \{da\}^{k} \{d(1-b)\}^{p-k} \nonumber \\ 
 + \{da\}^{p-k} \{d(1-b)\}^{k} \left. \right] \times \int_{0}^{\infty} d \xi_{2} \int_{\xi_{2}}^{\infty} d \xi_{1} (\xi_{1}^{k} \xi_{2}^{p-k} + \xi_{1}^{p-k} \xi_{2}^{k} ) \Psi(\xi_{1}) \Psi(\xi_{2}) . \label{eq:asym_inequalities}
\end{eqnarray}
 Here, we assume that a scaled PDF obeys a power-law tail $\Psi(\xi) \sim 1 / \xi^{1+s}$, which is supported by the results of Monte Carlo simulations. It can be estimated that in the limit $p \to s$, two integral terms in the above inequalities (\ref{eq:asym_inequalities}) satisfy the next relations. 
\begin{eqnarray}
\int_{0}^{\infty} d \xi_{2} \int_{\xi_{2}}^{\infty} d \xi_{1} \xi_{1}^{p} \Psi(\xi_{1}) \Psi(\xi_{2}) \gg \int_{0}^{\infty} d \xi_{2} \int_{\xi_{2}}^{\infty} d \xi_{1} \xi_{2}^{p} \Psi(\xi_{1}) \Psi(\xi_{2}),  \\
\int_{0}^{\infty} d \xi_{2} \int_{\xi_{2}}^{\infty} d \xi_{1} \xi_{1}^{p} \Psi(\xi_{1}) \Psi(\xi_{2}) \simeq \mu_{p} \to \infty. 
\end{eqnarray}
 This immediately follows that the growth rate of the system $\gamma$ and power-law exponent $s$ of PDF are determined by the following transcendental equation. 
\begin{eqnarray}
s \gamma = \{c(1-a)\}^{s} + \{da\}^{s} - 1, \label{eq:asym_teq} \\
\gamma = \frac{1}{2} \left[ c(1-a+b) + d(1+a-b) \right] + \frac{1}{2} (c-d)(1-a-b) \ A - 1, \label{eq:asym_gamma} \\
A = \frac{1}{\mu_{1}} \int_{0}^{\infty} d \xi_{2} \int_{\xi_{2}}^{\infty} d \xi_{1} (\xi_{1} - \xi_{2}) \Psi(\xi_{1}) \Psi(\xi_{2}). 
\end{eqnarray}
 Consequently, we can successfully derive a transcendental equation (\ref{eq:asym_teq}) which determines the power-law exponent $s$ for $s > 1$. Note that the value of integral $A$ in Eq.~(\ref{eq:asym_gamma}) is unable to be calculated analytically, thus a numerical integration must be performed to solve the transcendental equation. Therefore, the possibility of the existence of a power-law tail in the PDF can be verified by this analytical caluculations. The exponent $0 \le s \le 1$ is also able to be obtained in the same manner as mentioned in Sec.~\ref{sec:keypoint}. It is found that when $s \to 1$, $A \to 1$\cite{rf:FOY2004} and therefore $\gamma$ and $s$ are connected continuously from $s > 1$ to $0 \le s \le 1$ and are continuous functions of the interaction parameters $a, b, c, d$. A typical example of the comparison between the theory and the result of Monte Carlo simulation is illustrated in Fig.~\ref{fig:asymmetry}(a). A very good agreement is achieved on the right side of PDF. As can be seen from Eq.~(\ref{eq:asym_gamma}), there have two cases (I) $c=d$ and (II) $a+b=1$ that $s$ and $\gamma$ can be derived without evaluating the integral $A$ numerically. It can be shown that these special cases have the following interesting features. In the case (I), the parameter $b$ is totally eliminated from the transcendental equation despite imposing no restrictions on $b$. This means that the exchange rate of smaller quantities $b$ is unimportant for the system and have no effect on $\gamma$ and $s$. In the case (II), the exponent $s$ is a continuous function of three interaction parameters $a, c, d$ and a phase diagram of $s$ fixed at $d = 0.5$ is illustrated in Fig.~\ref{fig:asymmetry}(b). It is shown that the diagram of AMI is asymmetrically distorted compared with that of SMI\cite{rf:bBLR2003}. In the hatched region in Fig.~\ref{fig:asymmetry}(b), a power-law tail disappears because of the divergence of the solution of transcendental equation. It is found that the region is unbounded in AMI, while it is bounded in SMI\cite{rf:FOY2004}. 

\section{Effects of randomness in interaction}

 Next, we investigate many-body stochastic processes with multiplicative interactions with randomly varying interaction parameters\cite{rf:OFY2004}. For simplicity, the form of interaction rule is chosen to be the one of SMI here, that is $x' = \alpha x + \beta y, \ y' = \beta x + \alpha y$. The difference from SMI is that at each timestep, the interaction parameters $\alpha, \beta$ are randomly determined by a given probability distribution $\rho(\alpha, \beta)$. As in the previous section, this model is also considered with following the procedures in Sec.~\ref{sec:keypoint}. A master equation for PDF is given by
\begin{eqnarray}
\frac{\partial }{\partial t} f(z, t) &=& \int_{0}^{\infty} d \alpha \int_{0}^{\infty} d \beta \ \rho(\alpha, \beta) \int_{0}^{\infty} dx \int_{0}^{\infty} dy \ f(x, t) f(y, t) \nonumber \\ 
				     & & \hspace{40mm} \times \left[ \delta(z - (\alpha x + \beta y)) - \delta(z - x) \right]. \label{eq:rand_mastereq}
\end{eqnarray}
 Using the same scaling relations (\ref{eq:scaling}), we have the scaled moment inequalities of this processes.  
\begin{eqnarray}
\frac{\sum_{k=1}^{k_p-1} (\overline{\alpha^{k} \beta^{p-k}} + \overline{\alpha^{p-k} \beta^{k}}) \mu_{k} \mu_{p-k} }{ p \mu_{p} - \left( \overline{\alpha^{p}} + \overline{\beta^{p}} - 1 \right)} \hspace{10mm} \nonumber \\
 \le \ \mu_{p} \ \le \ \frac{\sum_{k=1}^{k_p} (\overline{\alpha^{k} \beta^{p-k}} + \overline{\alpha^{p-k} \beta^{k}}) \mu_{k} \mu_{p-k} }{ p \mu_{p} - \left( \overline{\alpha^{p}} + \overline{\beta^{p}} - 1 \right)}, 
\end{eqnarray}
 where $\gamma = \overline{\alpha} + \overline{\beta} - 1$ and the overlines denote the averages with respect to $\rho(\alpha, \beta)$. These inequalities mean that the moment $\mu_p$ diverges at the point where the denominators of upper and lower bounds of $\mu_p$ are equal to zero. This indicates that the PDF obeys a power-law tail $\Psi(\xi) \sim 1/\xi^{1+s}$ whose exponent $s > 1$ is determined by the following transcendental equation.
\begin{equation}
s \gamma = \overline{\alpha^s} + \overline{\beta^s} - 1.
\end{equation}
 The exponent $0 \le s \le 1$ can be derived in the same manner as mentioned in Sec.~\ref{sec:keypoint}. Interesting results are found by comparing the growth rate $\gamma$ and power-law exponent $s$ of this model with those of SMI whose interaction parameters are given by $\overline{\alpha}$ and $\overline{\beta}$. The transcendental equation of this SMI is described by 
\begin{equation}
s' \gamma' = \overline{\alpha}^{s'} + \overline{\beta}^{s'} - 1,
\end{equation}
 where $\gamma'$ and $s'$ are the growth rate and power-law exponent of the SMI. From Jensen's inequality\cite{rf:Jensen}, it immediately follows that $\overline{\alpha^s} > \overline{\alpha}^s \ (s > 1)$ and $\overline{\alpha^s} < \overline{\alpha}^s \ (0 < s < 1)$. This concludes that 
\begin{equation}
\gamma < \gamma' \ (0 < s < 1), \hspace{5mm} s < s'. 
\end{equation}
 These inequalities indicate that randomness in interaction parameters have effects of reducing the values of both the growth rate and power-law exponent in general. In the next subsection, we give one concrete example to check the effects.

\subsection{Two-peaked system}

\begin{figure}
\begin{tabular}{cc}
\resizebox{60mm}{!}{\includegraphics{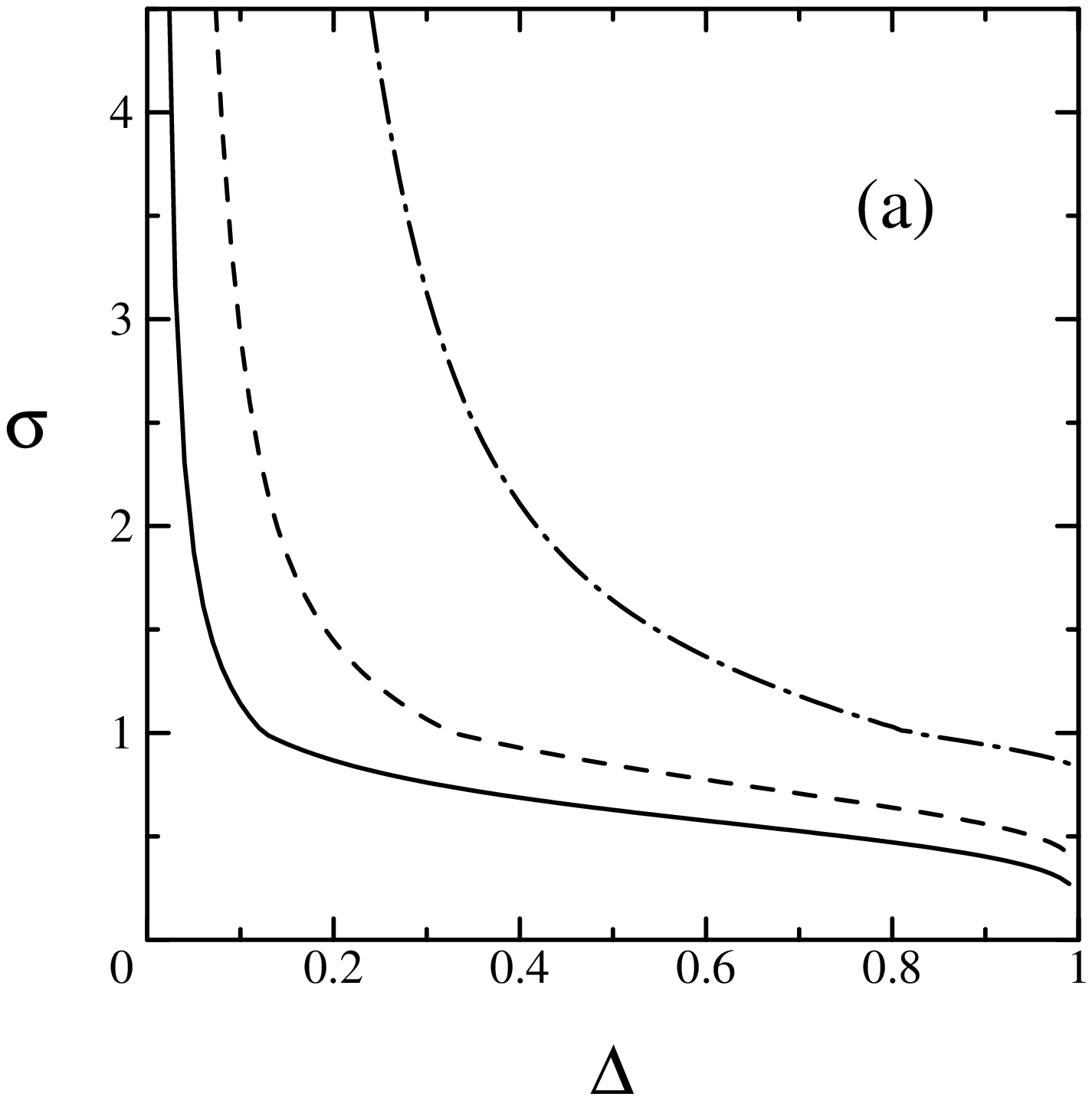}}
\resizebox{60mm}{!}{\includegraphics{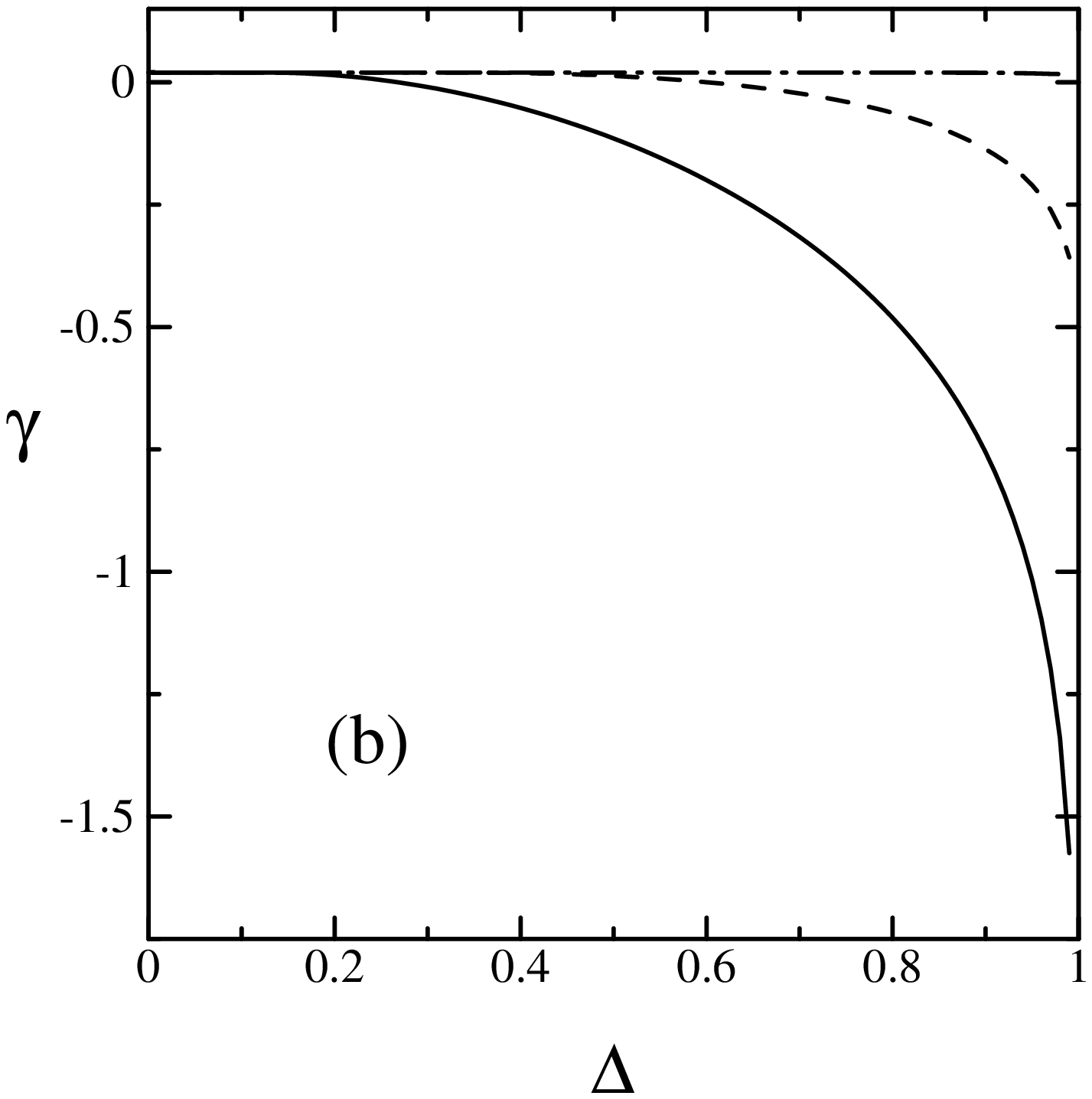}}
\end{tabular}
\caption{ Theoretically evaluated curves of (a) power-law exponents $\sigma$ and (b) growth rates $\gamma$ in two-peaked system at $p=0.9$ (solid lines), $p=0.5$ (dashed lines), $p=0.1$ (dash-dotted lines) with $\alpha_{0}=1.01$ and $b_{0}=0.01$. }
\label{fig:twopeaked}
\end{figure}

 In this example, $\rho(\alpha, \beta)$ is given by 
\begin{equation}
\rho(\alpha, \beta) = p \delta(\alpha - \alpha_1) \delta(\beta - \beta_1) + (1-p) \delta(\alpha - \alpha_2) \delta(\beta - \beta_2),
\end{equation}
 where $0 \le p \le 1$ and $\alpha_1, \beta_1, \alpha_2, \beta_2$ are certain positive constants. Here, for convenience of explanation, we define an index representing the displacement from SMI $\Delta$ as follows.
\begin{equation}
\Delta \equiv (1-p) (\alpha_2 - \alpha_1) / \alpha_0 = (1-p) (\beta_2 - \beta_1) / \beta_0,
\end{equation}
 where $\alpha_0 = p \alpha_1 + (1-p) \alpha_2, \ \beta_0 = p \beta_1 + (1-p) \beta_2$. The magnitude of $\Delta$ indicates the strength of randomness. The theoretical curves of growth rates and power-law exponents are illustrated in Figs.~\ref{fig:twopeaked}.
 It is obvious that all of their values decrease monotonically as $\Delta$ becomes large. It is shown in Fig.~\ref{fig:twopeaked}(a) that the exponents drastically decrease even though $\Delta$ slightly increases. Surprisingly, it can be seen in Fig.~\ref{fig:twopeaked}(b) that the growth rates reduce from positve to negative when $\Delta$ is large enough.

\section{Summary and discussions}

 In this paper, we reported two extended models of many-body stochastic processes with multiplicative interactions from the standpoints of effects of asymmetry and randomness in interaction. It was found by investigating moment inequalities that their PDFs have power laws at the tails in general. We successfully obtained the values of growth rates and power-law exponents in both models. 

 In econophysics, several reserchers\cite{rf:IKR1998, rf:OFY2004, rf:FOY2004, rf:Slanina2004} think that these processes are one of the possible candidates for explaining power-law tails observed in wealth distributions which are often called Pareto law\cite{rf:Pareto1897} or Zipf's law\cite{rf:Zipf1949}. In economic phenomena, there is a well-known concept called the Matthew effect, which is also said "the rich gets richer and the poor gets poorer." This effect can be interpreted as a potential asymmerty in wealth exchange interactions between transacters. Therefore, it is more natural to use AMI for explaining wealth distributions. Remember that in a special case of AMI, the growth rate and power-law exponent do not depend on the interaction parameter of small quantities. This result would imply that the shapes of wealth distributions can be determined by rich people and can be unaffected by poor people. On the other hand, everyone has own strategy to transact its wealth in real situations. This can be the origin of randomness in interaction. Therefore, it is also more natural to consider effects of interaction randomness in the processes. As can be seen from the example of two-peaked system, only a few amount of particles that has large quantities is sufficient to decrease the values of the growth rate and power-law exponent enormously. This result would imply that the entry of a few very rich people into a economic system ends up with drastically reducing the economic growth rate and magnifying the degree of wealth inequality of the system. These implications seem to give us new perspectives on the analysis of wealth distributions.


\end{document}